\documentclass[]{aa}  
\usepackage{graphicx}
\usepackage{txfonts}
\begin{document}
   \title{The water ice rich surface of (145453) 2005 RR$_{43}$: \\a case for a 
   carbon-depleted population of TNOs?}

   \subtitle{}

   \author{N. Pinilla-Alonso
          \inst{1}
	  \and
	  J. Licandro
	  \inst{2,3}
	  \and          
	  R. Gil-Hutton
	  \inst{4}
	  \and          
	  R. Brunetto
	  \inst{5,6}
	  }

   \offprints{N. Pinilla-Alonso}

  \institute{Fundaci\'on Galileo Galilei \& Telescopio Nazionale Galileo, P.O.Box 565, E-38700, S/C de La Palma, Tenerife, Spain.\\
              \email{npinilla@tng.iac.es}
         	\and
	     Isaac Newton Group, E-38700, Santa Cruz de La Palma, Tenerife, Spain.\\
         	\and
	     Instituto de Astrof\'{\i}sica de Canarias, c/V\'{\i}a L\'actea s/n, E38205, La Laguna, Tenerife, Spain.\\
         	\and
	     Complejo Astron\'omico El Leoncito (Casleo) and San Juan National University, Av. Espa\~na 1512 sur,J5402DSP, San Juan, Argentina.\\
         	\and
	     Dipartimento di Fisica, Universit\`a del Salento, Via Arnesano, I-73100, Lecce, Italy.\\
                 \and
             INAF-Osservatorio Astrofisico di Catania, Via S. Sofia 78, I-95123, Catania, Italy.}
   \date{Submitted, February, 2007; accepted, April, 2007}


  \abstract
   {Recent results suggest that there is a group of trans-Neptunian objects 
   (TNOs) (2003 EL$_{61}$ being the biggest member), with surfaces composed 
   of almost pure water ice and with very similar orbital elements. These 
   objects provide exciting laboratories for the study of the processes that 
   prevent the formation of an evolved mantle of organics on the surfaces of 
   the bodies in the trans-Neptunian belt (TNb).}
   {We study the surface composition of another TNO that moves in a similar 
   orbit, (145453) 2005 RR$_{43}$, and compare it with the surface composition 
   of the other members of the group.}
   {We report visible and near-infrared spectra in the 0.53-2.4$\mu$m  
   spectral range, obtained with the 4.2m William Herschel Telescope and 
   the 3.58m Telescopio Nazionale Galileo at the ``Roque de los Muchachos" 
   Observatory (La Palma, Spain). Scattering models are used to derive 
   information about its surface composition. We also measure the depth $D$ 
   of the water ice absorption bands and compare with those of the other 
   members of the group.}
   {The spectrum of 2005 RR$_{43}$ is neutral in color in the visible and 
   dominated by very deep water ice absorption bands in the near infrared 
   ($D$= 70.3$\pm$2.1\% and 82.8 $\pm$4.9\% at 1.5 $\mu$m and 2.0 $\mu$m 
   respectively). It is very similar to the spectrum of the group of TNOs 
   already mentioned. All of them present much deeper water ice absorption 
   bands ($D >$40 \%) than any other TNO except Charon. Scattering models show 
   that its surface is covered by water ice, a significant fraction in 
   crystalline state with no trace (5\% upper limit)  of complex organics. 
   Possible scenarios to explain the existence of this population of TNOs 
   are discussed: a giant collision, an originally carbon depleted composition,
    or a common process of continuous resurfacing.}
   {2005 RR$_{43}$ is member of a group, may be a population, of TNOs 
   clustered in the space of orbital parameters that show abundant water ice 
   and no signs of complex organics and which origin needs to be further 
   investigated. The lack of complex organics in their surfaces suggests 
   a significant smaller fraction of carbonaceous volatiles like CH$_4$ 
   in this population than in "normal" TNOs. A carbon depleted population of TNOs could be the origin of 
   the population of carbon depleted Jupiter family comets already noticed by A'Hearn et al. (1995).}
   \keywords{   Kuiper Belt --
		Solar system : formation}
   \titlerunning{ The surface of (145453) 2005 RR43: a case for a carbon-depleted population of TNOs ?}
   \maketitle
%

\section{Introduction}

Spectroscopic and spectrophotometric studies show that about 70\% of TNOs 
present a mantle of complex organics on their surfaces ( Brunetto et al. 
\cite{Brunetto2006}). 
Long term processing by high energy particles and solar radiation on icy 
bodies, induces the formation of organic species in their outer layers, 
resulting in a mantle that covers the unprocessed original ices (Moore et 
al. \cite{Mooreetal83}; Johnson et al. \cite{Johnetal84}; Strazzulla \& 
Johnson \cite{StrazzullaJo91}). Until recently, the only case of a TNO with a 
surface covered basically by a thick layer of water ice was Charon (Buie 
et al.\cite{Buieetal1987}, Marcialis et al. \cite{Marcialisetal1987}), and it 
has been considered an intringuing case because of the need of a resurfacing 
mechanism like cryovolcanism or collisons with micro-meteorites 
(Brown \cite{Brown2002}; Cruikshank \cite{Cruik1998}).
Recently, it has been showed that (55636) 2002 TX$_{300}$ (Licandro et 
al. \cite{LicTX}) and (13308) 1996 TO$_{66}$ (Brown et al. \cite{BrownTO66}) 
also have surface composition similar to Charon. During last year the 
spectra of five other objects were published, revealing that their surfaces 
are also covered by fresh water ice: (136108) 2003 EL$_{61}$ (Trujillo et 
al. \cite{TrujEL61}), its biggest satellite  S/2005 (136108) 1 (Barkume et 
al. \cite{BarSatellite}) and during the review process of this paper 2003
OP$_{32}$, 1995 SM$_{55}$ and 2005 RR$_{43}$ (Brown et al. \cite{BrownFamily}).
The spectra of these TNOs show all the same characteristics, they are neutral 
and featureless in the visible and show strong water ice absorption bands in 
the near infrared. All these TNOs, except Charon, are located 
in a narrow region of the orbital parameters space (41.6 $<$a$<$ 43.6 AU, 25.8 
$<$i$<$ 28.2 deg., 0.10 $<$e$<$ 0.19). The existence of a population of TNOs with 
Charon-like surfaces and similiar orbital parameters needs to be explained 
as it can have a strong impact in the knowledge of the trans-neptunian belt 
formation theories and/or resurfacing mechanisms. 

In this paper we present visible and near-infrared spectroscopy of a member of 
this group, (145453) 2005 RR$_{43}$ (a,e,i = 43.06AU, 
0.14, 28.54 deg.) and derive compositional information using scattering models.
Finally we describe different scenarios that can explain the existence of this 
population of TNOs.

\section{Observations and analysis of the spectrum}

We obtained the visible spectrum of 2005 RR$_{43}$ with the 4.2m William Herschel telescope (WHT) and the near-infrared spectrum with the 3.58m "Telescopio Nazionale Galileo" (TNG) both  at the "Roque de los Muchachos" Observatory (Canary Islands, Spain).


The visible spectrum (0.35-0.95$\mu$m) was obtained on 2006 Sep. 23.18 UT 
using the low resolution grating (R158R, with a dispersion of 1.63 $\AA$/pixel) 
of the double-armed spectrograph ISIS at WHT, 
and a 4" slit width oriented at the parallactic angle. The tracking was at 
the TNO proper motion.

Three 900s exposure time spectra were obtained by shifting the object by 10" 
in the slit to better correct the fringing. Calibration and extraction of the 
spectra were done using IRAF and following  standard procedures 
(Massey et al. \cite{Massey1992}). TNO spectra were averaged and the reflectance 
spectrum was obtained by dividing the spectrum of the TNO by the spectrum of 
the solar analogue star Hyades 64 obtained the same night just before and 
after the observation of the TNO at a similar airmass. Final spectrum presented 
in Fig. 1 was smoothed using a smoothing box-car of 15 pixels to improve the 
S/N.


The near-infrared spectrum was obtained on 2006 Sep. 29.15 UT, using the low 
resolution spectroscopic mode of NICS (Near-Infrared Camera and Spectrometer) 
at the TNG based on an Amici prism disperser. This mode yields a complete 
0.8-2.4 $\mu$m spectrum. A 1.5" slit width corresponding to a spectral 
resolving  power $R\simeq34$ quasi-constant along the spectrum was used. The 
observing and reduction procedures were as described in Licandro et al. 
(\cite{Lic2002}). The total "on object" exposure time was 11160s.

To correct for telluric absorption and to obtain the relative reflectance, 
solar analogue star Hyades 64 and two G-2 Landolt stars (115-271 and 98-978, 
Landolt \cite{Landolt}) were observed at different airmasses during the night. 
The reflectance spectrum of the TNO was obtained using all the SA stars, 
averaged and then normalized to fit the visible one using the overlapping 
spectral region between 0.85-0.95 $\mu$m. The combined visible and 
near-infrared (VNIR) spectrum is presented in Fig. 1 
together with the spectra of the other members of the group.

\begin{figure}
	\centering
	\includegraphics[width=\columnwidth]{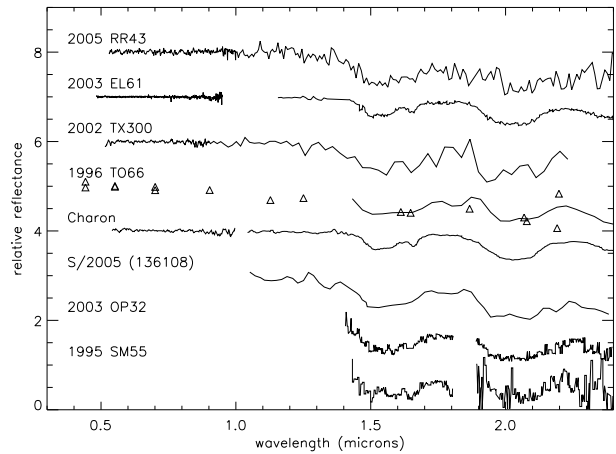}
	\caption{ Spectrum of 2005 RR$_{43}$, normalized at 0.55 $\mu$m, 
	together with the spectra of the other members of the group 
	shifted in the vertical axis for clarity. References in table 1.}
 	\label{Fig1}
 \end{figure}


The VNIR spectrum reveals three important characteristics: 
(a) The visible is featureless within the S/N. There is no clear evidence of 
any absorption reported for other TNOs (e.g Lazzarin et al. \cite{Lazzarin}; 
Fornasier et al. \cite{Fornetal2004a}); (b) It is neutral (the spectral slope, 
computed between 0.53 and 1.00 $\mu$m, is $S'=0.4\pm1\%/1000\AA$), far from 
the more characteristical red slopes in the TNb (Fornasier et 
al. \cite{Fornetal2004b}); (c) It presents two deep absorption bands centered at 1.5 and 2.0 $\mu$m, indicative of water ice. 

The flat visible spectrum is indicative of the lack (or very low abundance) of complex organic and/or silicates in the surface of this TNO. 

The depth of the water ice bands is a good indicator of its abundance on the surface of icy objects, it is also a reasonably good marker of the size of the icy particles and of the contamination by non-ice components (Clark et al.\cite{Clarkbands}). We computed the depth of the bands, \textit{D}, with respect to the continuum of the spectrum as \textit{D=1-R$_{b}$/R$_{c}$}, where \textit{R$_{b}$} is the reflectance in the center of the band and \textit{R$_{c}$} is the reflectance of the continuum at 1.2 $\mu$m. For 2005 RR$_{43}$, $D$ is 70.3$\pm$2.1\% and 82.8 $\pm$4.9\% at 1.5 $\mu$m and 2.0 $\mu$m respectively, beeing the deepest water ice absorption bands ever observed in a TNO. 

Thus, we conclude that the surface of 2005 RR$_{43}$ is composed by a large fraction of large sized (or a thick layer of) water ice particles, and none or a very small fraction of complex organics or silicates.


This is confirmed by scattering models. We use the simple
one-dimensional geometrical-optics formulation 
by Shkuratov et al. (\cite{Shkuratov}), to obtain information about the 
surface composition. 

It is important to determine if water ice is in amorphous or crystalline state as it can be indicative of resurfacing processes. Crystalline water ice can be easily identified by an absorption band at 1.65 $\mu$m. This band has been detected in the spectrum of TNOs Charon (Brown \& Calvin \cite{BrownCharon}) and 2003 EL$_{61}$ (Trujillo et al. \cite{TrujEL61}), the only TNOs that have spectra similar to 2005 RR$_{43}$ and sufficiently high signal-to-noise to see this band. Unfortunately, the relatively poor S/N of our spectrum does not allow us to clearly detect this feature. 

So we first tried to model 2005 RR$_{43}$ spectrum with pure amorphous water ice, but all the tests we made produced a band at 1.5 $\mu m$ narrower than that seen in the spectrum (see Fig. 2). We then tried with pure crystalline water ice obtaining a better fit. Finally we used an intimate mixture of crystalline and amorphous water ices obtaining even better results and improving the fit above 2 $\mu$m, although this part of the spectrum is too noisy to be a significant constraint. We thus conclude that water ice in the surface of 2005 RR$_{43}$ is, at least in a significant fraction, crystalline.

On the other hand, all these models allow a small percentage of minor components such as amorphous carbons or silicates (e.g. Olivine), up to 5\%. We notice that the S/N of the spectrum and the model itself do not allow us to make a detailed study of the surface mineralogy. An important parameter to include, that could help to better constraint the models and, in particular, determine the abundance of minor dark constituents, is the albedo of the TNO, as even a small amount of them can darken the surface significantly. 

\begin{figure}
	\centering
	\includegraphics[width=\columnwidth]{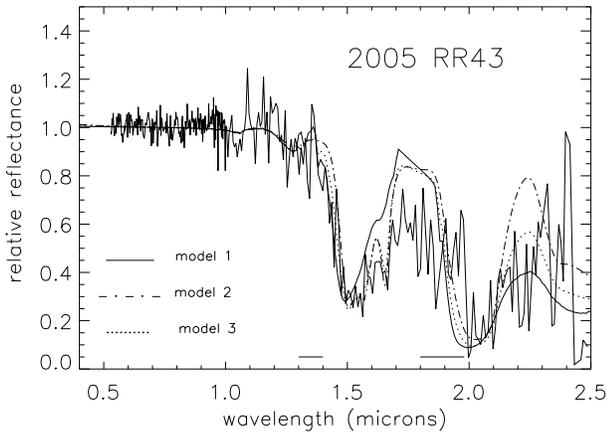}
	\caption{Best fits of the spectrum with the Shkuratov approximation (intimate mixtures of amorphous water ice (AWI), crystalline water ice (CWI) and olivine (O)). Model 1: 95\% AWI (70 $\mu$m) + 5\% O (30 $\mu$m) ; model 2: 95\% CWI (50 $\mu$m) + 5\% O (30 $\mu$m); model 3: 55\% AWI (60 $\mu$m) + 40\% CWI (70 $\mu$m) + 5\% O (30 $\mu$m). The width of 1.5 $\mu$m band is best fitted with crystalline water ice.}
	\label{Fig2}
\end{figure}

\section{Discussion and conclusions}

To date, about 40 TNOs have been observed spectroscopically in the 
near-infrared. Among them only seven exhibit a spectrum similar to that of 
2005 RR$_{43}$. Table 1 summarizes their 1.5 $\mu$m band depth (\textit{D}) 
and spectral slope $S'$ computed as in Sec. 3 for 2005 RR$_{43}$. 
Notice that all of these TNOs have 
very deep water ice absorptions, $D > 40$\% {\ and $S'\sim 0$.}

\begin{table}
\centering
\begin{tabular}{l c c c l}
\hline
\hline
object & H$_V$ & \textit{S'} & \textit{D} & Ref \\  \hline \hline
Charon & 0.9 & -0.6 & 41.5$\pm$3.4\% & (1,2)\\ \hline
1995 SM$_{55}$ & 4.8 & 2.4& 70.2$\pm$5.9\% & (3,4)\\ \hline
1996 TO$_{66}$ & 4.5 & 4.5 & 59.1$\pm$1.9\% & (3,5,6)\\ \hline 
2002 TX$_{300}$ & 3.3 & 1& 64.1$\pm$3.9\% & (7)\\ \hline 
2003 OP$_{32}$ & 4.1 & -1.1& 64.2$\pm$2.5\% & (4,8)\\ \hline
(136108) 2003 EL$_{61}$ & 0.2 & -0.4& 40.7$\pm$ 2.1\% & (8,9) \\ \hline
S/2005 (136108) 1 & 3.5 & NA & 69.1$\pm$1.0 \% & (10) \\ \hline
2005 RR$_{43}$ & 4.0 &  -0.4 &70.3$\pm$2.1 \% & (11)\\ \hline
\end{tabular} 

\caption{H: absolute magnitud; \textit{S'}: spectral gradient; \textit{D}: 
absorption at 1.5 $\mu$m.
References: 1. Brown \& Calvin \cite{BrownCharon}, 
2. Fink \& DiSanti \cite{Fink}, 3. The online updated database from 
Hainaut \& Delsanti \cite{MBOSS}, 4. Brown et al. \cite{BrownFamily}, 
5. Brown et al.\cite{BrownTO66}, 6. Noll et al.\cite{NollTO66}, 
7. Licandro et al.\cite{LicTX}, 8. Tegler et al.\cite{TegEL61}, 
9. Trujillo et al \cite{TrujEL61}, 
10. Barkume et al.\cite{BarSatellite}, 11. This work.
}
\end{table}

Brown et al. (\cite{BrownFamily})
noticed that, excluding Charon, these objects present very similar spectral
properties and orbital parameters (see Fig. 2 of Brown's paper). 
On the other hand, the spectra and colors of other TNOs in the neighbourhood 
of the cluster, e.g. (20000) Varuna (Licandro et al.  \cite{LicVaruna}) or 
(50000) Quaoar (Jewitt \& Luu \cite{JewittQuaoar}), are different from the 
spectrum of 2005 RR$_{43}$ and present the variety of colors and composition 
observed in other regions of the TNb. Thus, we also conclude that objects in 
Table 1 (except Charon) are part of a population clustered in the space of 
parameters and with similar surface properties different from those of the 
objects in the neighbourhood.

There are other smaller TNOs, with no photometric or spectroscopic data 
published, that according to their orbital parameters, could be also members 
of this population like 1995 GJ, 1999 OK$_{4}$, 1999 RA$_{215}$, 2003 FB$_{130}$, 
2005 PM$_{21}$. Photometric and/or spectroscopic observations are needed.


Water ice surfaces with no traces of organics should not be common in the TNb
assuming that the original chemical composition of all TNOs is very similar: 
objects composed of abundant water ice, some molecular ices like CO, CO$_2$, 
CH$_4$,N$_2$ and silicates.  Long term processing by high energy particles and 
solar radiation induces the formation of complex organic species in the outer 
layers of the TNOs resulting in a dark and usually red mantle that covers the 
unprocessed original ices. Thus, the existence of a population of TNOs with 
no signs of organics, in a very narrow region of the space of orbital 
parameters is an intriguing fact that needs to be explained. 
We discuss three different scenarios that need to be further studied:

1) {\bf Collisional family}: 
The destruction of the irradiation mantle by an 
energetic collision has been proposed by Licandro et al. (\cite{LicTX}) to 
explain the fresh surface of one of the member of this population, 
2002 TX$_{300}$. Such an impact breaks the 
irradiation mantle and produces enough energy to sublimate a certain amount 
of ices on the upper layers of the body (Gil-Hutton \cite{Gil-Hutton}). 
The material sublimated by the collision can be globally redeposited over the 
TNO on a timescale of tens of hours (Stern \cite{Stern}) and the low vertical 
diffusion velocity of the ice ensures that large particles can be formed with 
a high efficiency while being downward transported and deposited on the 
surface. 
Brown et al. (\cite{BrownFamily}) proposed that the TNOs listed 
in Table 1 (but Charon) are fragments produced by a catastrophic 
collision suffered by 2003 EL$_{61}$. This would explain the rapid rotation of 
this large object, the existence of two satellites orbiting it, and the clustering
of objects in a small region of the TNb with similar surface properties.
 
Although this scenario appears promising, it has several problems: 
First, the low mean intrinsic collisional 
probability makes highly improbable a catastrophic collision with a large 
projectile capable of providing enough kinetic energy to disperse large 
fragments to their present positions, even in a TNb more massive that the 
present one. 
Second, it is difficult to model the collisional process to 
produce such a family working with so few members and given the large 
uncertainties in their orbital elements. 
Third, the current dispersion in orbital elements of the fragments shows orbital
elements expected from a too high dispersive velocity of 400 m/s. Brown et al.
argued that only 2003 EL$_{61}$ does not fit withing a smaller velocity of 140
m/s and argued that this TNO has large excursions in eccentricity over time
owing to chaotic diffusion within the 12:7 mean-motion resonance (MMR) with Neptune. 
If this is the case, following the analysis by Nesvorny and Roig (\cite{NesRoig01}) of the 12:7 MMR with
Neptune, and cited by Brown et al. (\cite{BrownFamily}), the chaotic diffusion enlarge
the initial eccentricity of any orbit near its borders in such a way that
it becomes Neptune grazing and the body escapes in less than $10^8\; yr$,
dispersing very fast any object in its vicinity.
But the collision must have happened in the distant past,
when the belt was far more crowded with large objects (Morbidelli
\cite{Morbi2007}).
Fourth, if the collision happened in the distant past, the effects of 
long term processing by high energy particles and solar radiation should be
present and are not. Gil-Hutton (\cite{Gil-Hutton}) shows that the time scale 
required to form a black irradiation mantle of carbon residues in a TNO 
is $\sim$ 6 x 10$^{8}$ years. The competition of this effect with 
resurfacing due to high frequency 
impacts will increase the time scale to 10$^{9}$ years while intermediate 
states would result in red spectra. Moreover the efficiency of this process 
depends on the presence of organics on the surface so the absence of an 
irradiation mantle can  be explained by the loss of organics or the absence 
of them in the original composition. Thus, a possible solution of this problem
is that the loss of organics was produced by the collision and recondensation
process that formed the family.

2) {\bf Originally Carbon Depleted Population}:
 Another possible scenario is a population of objects 
originally carbon depleted, strongly concentrated in the space of orbital 
parameters, and in a region that it is dynamically unstable 
(the hot population). But this has also several problems: 
why carbon depleted objects 
were formed in a narrow region of the solar nebula and remained grouped until 
present time?. However, the existence of a population of carbon depleted TNOs
is an interesting case, as it could be the origin of the population of carbon 
depleted Jupiter family comets already noticed by A'Hearn et al. (\cite{Ahearn}) 
and whose origin remains unknown.

3) {\bf Continuous resurfacing process}:
Another possible scenario is that these objects are exposed to a mechanism 
that replenishes their surfaces permanently with fresh material from their 
interiors like cryovolcanism. Anyhow, it is difficult to explain why this 
mechanism affects so efficiently this group of TNOs and only this group.


In conclusion, 
the spectrum of TNO 2005 RR$_{43}$ in the visible and near-infrared range shows 
that its surface is covered by large water ice grains. Scattering models reveal
that the observed water ice is, at least in a significant fraction, 
crystalline. 2005 RR$_{43}$ spectrum is very similar to those of TNOs Charon, 
1996 TO$_{66}$, 2002 TX$_{300}$, 2003 OP$_{32}$, 1995 SM$_{55}$,
(136108) 2003 EL$_{61}$ and 
S/2005 (136108). It also has orbital elements very similar to those of these 
last four TNOs, supporting the existence of a population of TNOs with their 
surface covered by fresh water ice and almost no complex organics.
The lack of complex organics suggests 
a signficant smaller fraction of carbonaceous volatiles like CH$_4$ 
in this population than in "normal" TNOs. 
Such carbon depleted population of TNOs could be the origin of 
the population of carbon depleted Jupiter family comets already noticed 
by A'Hearn et al. (1995). The origin of this population needs to be 
further investigated.

\begin{acknowledgements}
We wish to thank Ted Roush for providing optical constants and Humberto Campins for his usefull comments.

\end{acknowledgements}

\end{document}